\documentclass[letter]{aa} % for the letters 
\usepackage{graphicx}
\usepackage{txfonts}
\usepackage{natbib}
\bibpunct{(}{)}{;}{f}{}{,} % to follow the A&A style

\newcommand{\kms}{\,$\mathrm{km\, s^{-1}}$}
\newcommand{\ms}{\,$\mathrm{m\, s^{-1}}$}
\newcommand{\Teff}{\ensuremath{T_{\mathrm{eff}}}}
\newcommand{\logg}{\ensuremath{\log g}}
\newcommand{\moh}{\ensuremath{[\mathrm{Fe/H}]}}

\newcommand{\Mj}{\ensuremath{M_{\mathrm{Jup}}}}

\begin{document}

  \title{Three planetary companions around M67 stars.\thanks{Based on observations collected at the ESO 3.6m telescope (La Silla), 
   at the 1.93m telescope of the Observatoire de Haute-Provence (OHP) 
   and at the Hobby Eberly Telescope (HET). }}
       
  \author{A. Brucalassi\inst{1,2} \and  L. Pasquini\inst{3} \and R. Saglia\inst{1,2} \and M.T. Ruiz\inst{4} \and P. Bonifacio\inst{5} 
  \and L. R. Bedin\inst{6} \and K. Biazzo\inst{7} \and C. Melo \inst{8}\and \\ 
  C. Lovis\inst{9} \and S. Randich\inst{10}   }
    
   \institute{Max-Planck f\"ur extraterrestrische Physik, Garching bei M\"unchen, Germany
   \and University Observatory Munich, Ludwig Maximillian Universitaet, Scheinerstrasse 1, 81679 Munich, Germany
   \and ESO -- European Southern Observatory, Karl-Schwarzschild-Strasse 2, 85748 Garching bei M\"unchen, Germany 
   \and Astronomy Department, Universidad de Chile, Santiago, Chile 
   \and GEPI, Observatoire de Paris, CNRS, Univ. Paris Diderot, Place Jules Janssen 92190 Meudon, France
   \and Istituto Nazionale di Astrofisica, Osservatorio Astronomico di Padova, Padova, Italy
   \and Istituto Nazionale di Astrofisica, Osservatorio Astronomico di Catania, Catania, Italy
   \and ESO -- European Southern Observatory,Santiago, Chile
   \and Observatoire de Geneve, Sauverny, CH
   \and Istituto Nazionale di Astrofisica, Osservatorio Astrofisico di Arcetri, Firenze, Italy
   }          
             
   \date{Received September 1, 2013; accepted December 13, 2013}

   \abstract{
   For the past six years we have carried out a search for massive
   planets around main sequence and evolved stars in the open cluster
   (OC) M67, using radial velocity (RV) measurements obtained with
   HARPS at ESO (La Silla), SOPHIE at OHP and HRS at HET.  Additional
   RV data come from CORALIE at the Euler Swiss Telescope.  We aim to
   perform a long-term study on giant planet formation in open
   clusters and determine how it depends on stellar mass and chemical
   composition.\\
   We report the detection of three new extrasolar planets: two in
   orbit around the two G dwarfs YBP1194 and YBP1514, and one around
   the evolved star S364.  The orbital solution for YBP1194 yields a
   period of 6.9 days, an eccentricity of 0.24, and a minimum mass of
   0.34 \Mj.  YBP1514 shows periodic RV variations of 5.1 days, a
   minimum mass of 0.40 $\Mj$, and an eccentricity of 0.39. The best
   Keplerian solution for S364 yields a period of 121.7 days, an
   eccentricity of 0.35 and a minimum mass of 1.54 \Mj.  An analysis
   of H$\alpha$ core flux measurements as well as of the line
   bisectors spans revealed no correlation with the RV periods,
   indicating that the RV variations are best explained by the
   presence of a planetary companion.  Remarkably, YBP1194 is one of
   the best solar twins identified so far, and YBP1194b is the first
   planet found around a solar twin that belongs to a stellar cluster.
   In contrast with early reports and in agreement with recent
   findings, our results show that massive planets around stars of
   open clusters are as frequent as those around field stars.}


   \maketitle
%
%________________________________________________________________

\section{Introduction}

In 2008 we began monitoring radial velocities (RVs) of a sample of
main sequence and giant stars in the open cluster (OC) M67, to detect
signatures of giant planets around their parent stars.  An overview of
the sample and of our first results is reported in
\citet{Pasquini2012}.  The goal of this campaign is to study the
formation of giant planets in OCs to understand whether a different
environment, such as a rich cluster like M67, might affect the planet
formation process, the frequency, and the evolution of planetary
systems with respect to field stars.  In addition, searching for
planets in OCs enables us to study the dependence of planet formation
on stellar mass and to compare the chemical composition of stars with
and without planets in detail.  Stars in OCs share age and chemical
composition \citep{Randich2005}, therefore it is possible to strictly
control the sample and to limit the space of parameters in a better
way than when studying field stars.
To address these questions we started a search for planets around
stars of the OC M67.  This cluster has solar age \citep[3.5-4.8
  $\mathrm{Gyr}$;][]{Yadav2008} and solar metallicity
\citep[$+$0.03$\pm$0.01 $\mathrm{dex}$;][]{Randich2006}.  In this
letter, we present the RV data obtained for the stars YBP1194,
YBP1514, and S364 that reveal the presence of Jovian-mass companions.

%__________________________________________________________________

\section{Stellar characteristics}

The three stars belong to the M67 sample presented in
\citet{Pasquini2012} with a proper motion membership probability
higher than 60\% according to \citet{Yadav2008} and
\citet{Sanders1977} (see online material).  The basic stellar
parameters ($V$, $B-V$, $\Teff$, $\logg$ and $\moh$) with their
uncertainties were adopted from the literature. Considering a distance
modulus of 9.63$\pm$0.05 \citep{Pasquini2008} and a reddening of
E(B-V)=0.041$\pm$0.004 \citep{Taylor2007}, stellar masses and radii
were estimated using the 4 Gyr theoretical isochrones from
\citet{Pietrinferni2004} and \citet{Girardi2000}.  The parameters
derived from isochrone fitting are comparable, within the errors, with
the values adopted from the literature. The main characteristics of
the three host stars are listed in Table 1.  We note that the errors
on these values do not include all potential systematics (see online
material). \\
\begin{table}
\caption{Stellar parameters of the three M67 stars hosting planets}
\label{StarParam}
\centering
\small
\begin{tabular}{lrrr}
\hline
\textbf{Parameters}&YBP1194&YBP1514&SAND364\\
\hline
$\mathrm{\alpha}$ $(\mathrm{J2000})$&  08:51:00.81 &  08:51:00.77 &  08:49:56.82 \\
$\mathrm{\delta}$ $(\mathrm{J2000})$& +11:48:52.76 & +11:53:11.51 & +11:41:33.00 \\
Spec.type       &G5V    &G5V& K3III\\
\ensuremath{m_{\mathrm{V}}}           $[\mathrm{mag}]$       &14.6\tablefootmark{a}           &14.77\tablefootmark{a}         &9.8\tablefootmark{b} \\
\ensuremath{B-V}                $[\mathrm{mag}]$       &0.626\tablefootmark{a}          &0.680\tablefootmark{a}         &1.360\tablefootmark{b}\\
\ensuremath{M\star}          [\ensuremath{M_{\odot}}]   &1.01$\pm$0.02\tablefootmark{c}  &0.96$\pm$0.01\tablefootmark{d} &1.35$\pm$0.05\tablefootmark{d}\\
\ensuremath{R\star}          [\ensuremath{R_{\odot}}]   &0.99$\pm$0.02\tablefootmark{d}  &0.89$\pm$0.02\tablefootmark{d} &21.8$\pm$0.7\tablefootmark{d}\\
$\logg$            $[\mathrm{cgs}]$       &4.44$\pm$0.035\tablefootmark{c} &4.57$\pm$0.05\tablefootmark{e} &2.20$\pm$0.06\tablefootmark{f}\\
$\Teff$            $[\mathrm{K}]$         & 5780$\pm$27\tablefootmark{c}   &5725$\pm$45\tablefootmark{e}   &4284$\pm$9\tablefootmark{f}\\
$\moh$           $[\mathrm{dex}]$       &0.023$\pm$0.015\tablefootmark{c}&0.03$\pm$0.05\tablefootmark{e} &$-$0.02$\pm$0.04\tablefootmark{f}\\
\hline
\end{tabular}
\normalsize 
\tablefoot{\tablefoottext{a}{\citet{Yadav2008}.}
\tablefoottext{b}{\citet{Montgomery1993}.}
\tablefoottext{c}{\citet{Onehag2011}.}
\tablefoottext{d}{\citet{Pietrinferni2004} and \citet{Girardi2000}.}
\tablefoottext{e}{\citet{Smolinski2011} and \citet{Lee2008}. }
\tablefoottext{f}{\citet{Wu2011}.}
}
\end{table}
YBP1194 is a G5V star, described by \citet{Pasquini2008} as one of the
five best solar analogs in their sample.  A detailed spectroscopic
analysis \citep{Onehag2011} has confirmed the star as one of the
best-known solar-twins.\\
YBP1514 also is a G5 main sequence star. 
We adopted the atmospheric parameters obtained by
\citet{Smolinski2011}, who used spectroscopic and photometric data
from the original Sloan Digital Sky Survey (SDSS-I) and its first
extension (SDSS-II/SEGUE).
These values are consistent, within the errors, with what has been
found in previous work on the same data by \citet{Lee2008} and in the
study of \citet{Pasquini2008}.\\
S364 (MMJ6470) is an evolved K3 giant star. 
The stellar parameters, summarized in Table \ref{StarParam}, are taken from \citet{Wu2011} . 
We derived its mass and radius by isochrone fitting \citep{Pietrinferni2004}.

\section{Radial velocities and orbital solutions}

The RV measurements were obtained using the HARPS spectrograph
\citep{Mayor2003} at the ESO 3.6m telescope in high-efficiency mode (
with R=90000 and a spectral range of 378-691 nm); the SOPHIE
spectrograph \citep{Bouchy2006} at the OHP 1.93 m telescope in
high-efficiency mode (with R=40000 and a spectral range of 387-694
nm), and the HRS spectrograph \citep{Tull1998} at the Hobby Eberly
Telescope (with R=60000 and a wavelength range of 407.6-787.5 nm).  In
addition, we gathered RV data points for giant stars observed between
2003 and 2005 \citep{Lovis2007} with the CORALIE spectrograph at the
1.2 m Euler Swiss telescope.  HARPS and SOPHIE are provided with a
similar automatic pipeline to extract the spectra from the detector
images and to cross-correlate them with a G2-type mask obtained from
the Sun spectra. Radial velocities are derived by fitting each
resulting cross-correlation function (CCF) with a Gaussian
\citep{Baranne1996,Pepe2002}.  For the HRS, the radial velocities were
computed using a series of dedicated routines based on IRAF and
cross-correlating the spectra with a G2 star template
\citep{Cappetta2012}.  All the observations for each star were
corrected to the zero point of HARPS, as explained in
\citet{Pasquini2012}, and were analyzed together.  Two additional
corrections were applied to the SOPHIE data, to take into account the
modification of the fiber link in June 2011 \citep{Perruchot2011} and
the low S/N ratio of the observations. For the first, we calculated
the offset between RV values of our stellar standard (HD32923) before
and after the change of the optical setup.  For the second, we
corrected our spectra using eq.(1) in \citet{Santerne2012}.  We
studied the RV variations of our target stars by computing the
Lomb-Scargle periodogram \citep{Scargle1982,Horne1986} and by using a
Levenberg-Marquardt analysis \citep[RVLIN]{Wright2009} to fit
Keplerian orbits to the radial velocity data.  The orbital solutions
were independently checked using the Yorbit program (Segransan et
al. 2013 in prep.).  For each case we verified that the RVs did not
correlate with the bisector span of the CCF (calculated following
\citet{Queloz2001}) or with the FWHM of the CCF.
\begin{table}
\caption{Orbital parameters of the planetary companions. \ensuremath{P}: period, \ensuremath{T}: time at periastron passage,
 \ensuremath{e}: eccentricity, $\omega$: argument of periastron, \ensuremath{K}: semi-amplitude of RV curve, 
 \ensuremath{m\sin{i}}: planetary minimum mass,
 $\gamma$: average radial velocity, $\sigma$(O-C): dispersion of Keplerian fit residuals.  }
\label{PlanetParam}
\centering
\small
\vspace{-5pt}
\begin{tabular}{lrrr}
%\hline\hline
\hline
\textbf{Parameters} &YBP1194&YBP1514&SAND364\\
\hline
\ensuremath{P}        $[\mathrm{days}]$& 6.958$\pm$0.001    & 5.118$\pm$0.001     & 121.710$\pm$0.305 \\
\ensuremath{T}        $[\mathrm{JD}]$  & 2455978.8$\pm$0.5  & 2455986.3$\pm$0.3   & 2456240.9$\pm$3.7    \\
\ensuremath{e}                         & 0.24$\pm$0.08      & 0.39$\pm$0.17       & 0.35$\pm$0.08     \\ 
$\omega$              $[\mathrm{deg}]$ & 98.62$\pm$25.68    & 327.49$\pm$16.05    & 273.51$\pm$12.81 \\
\ensuremath{K}        [\ms]            & 37.72$\pm$4.27     & 52.29 $\pm$10.39    & 67.42$\pm$5.85\\
\ensuremath{m\sin{i}} [\Mj]            & 0.34$\pm$0.05      & 0.40$\pm$0.11       & 1.54$\pm$0.24          \\
$\gamma$              [\kms]           & 34.184$\pm$0.006   & 34.057$\pm$0.017    & 33.217$\pm$0.018\\
$\sigma$(O-C)         [\ms]            & 11.55              & 14.6                & 15.0    \\
\hline
\end{tabular}
\normalsize

\end{table}
All the RV data for each star are available in the online material.\\
\\
YBP1194\\
We have acquired 23 RV measurements since 2008.  Fifteen were obtained
with HARPS with a typical S/N of 10 (per pixel at 550 nm), leading to
a mean measurement uncertainty of 13 \ms including calibration errors.
Eight additional RV measurements were obtained with SOPHIE and HRS
with mean measurement uncertainties of 9.0 \ms and 26.0 \ms.
%_____________________________________________________________
%                 A figure as large as the width of the column
%-------------------------------------------------------------
   \begin{figure}
   \vspace{-5pt}
   \centering
    \hspace{-16pt}
    \includegraphics[width=0.46\textwidth,height=3.5cm]{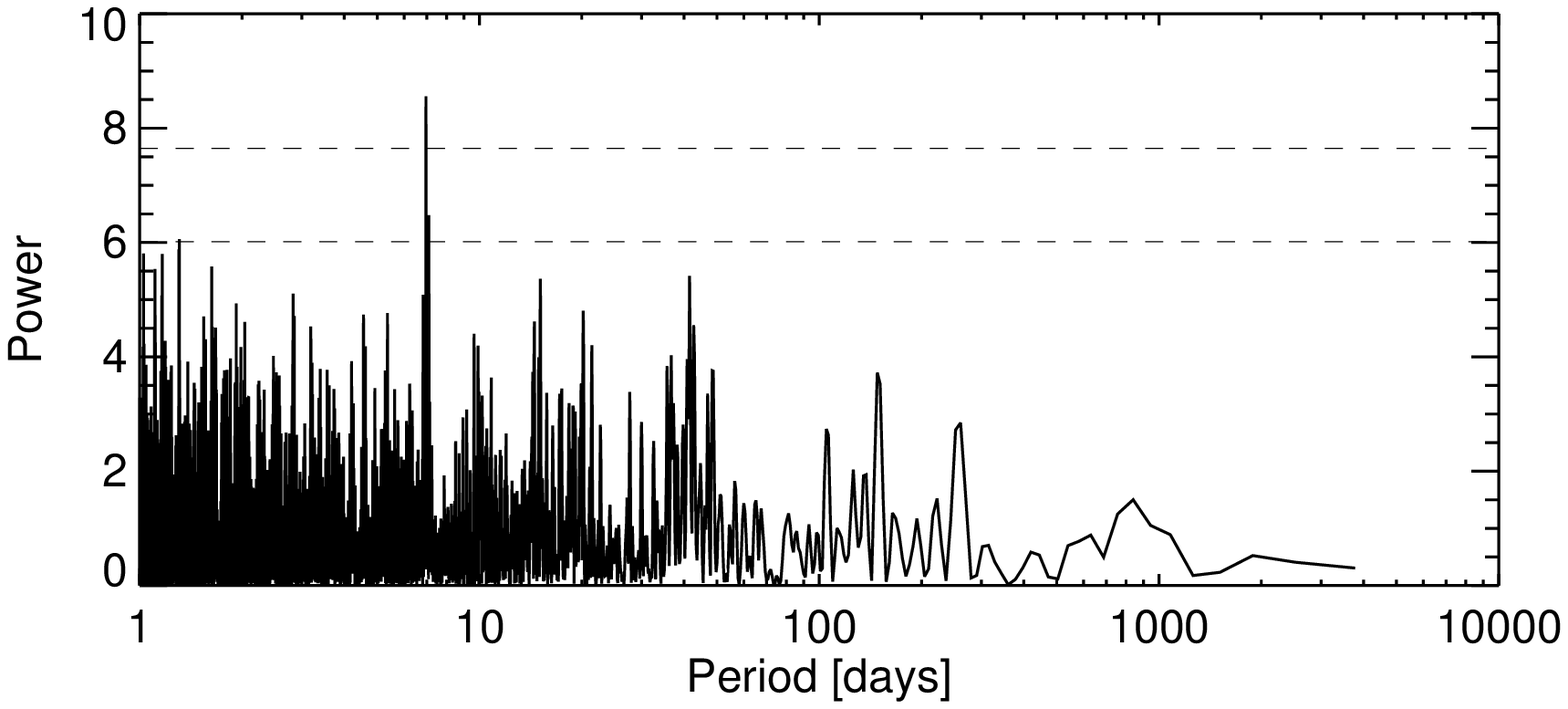}
    \includegraphics[width=0.48\textwidth]{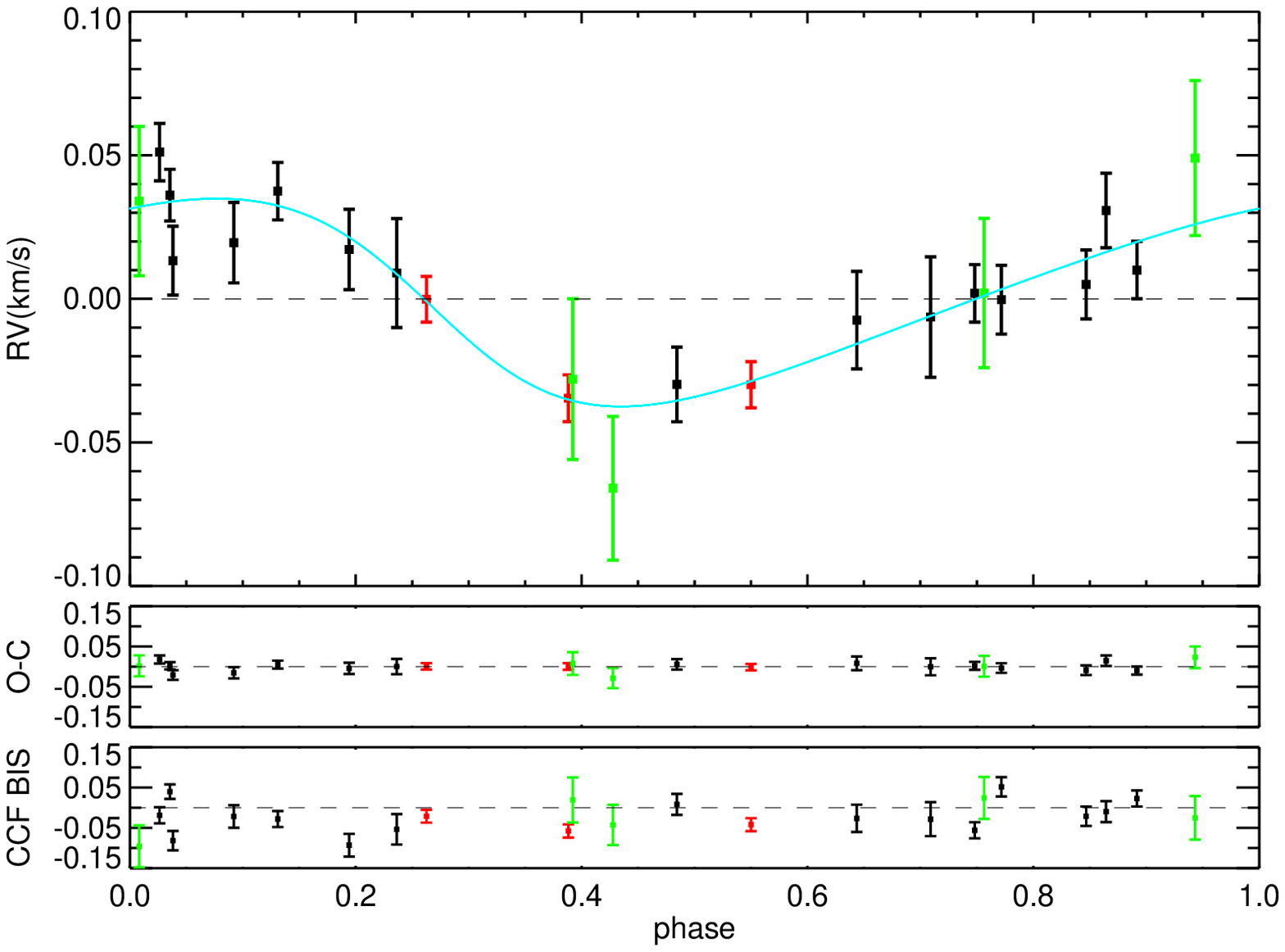}
      \vspace{-5pt}
      \caption{Top: Lomb-Scargle periodogram for YBP1194. 
      The dashed lines correspond to 5\% and 1\% false-alarm probabilities,
      calculated according to \citet{Horne1986} and white noise simulations.
      Bottom: phased RV measurements and Keplerian best fit, best-fit residuals, and 
      bisector variation for YBP1194.
      Black dots: HARPS measurements, red dots: SOPHIE measurements, green dots: HRS measurements.}
      \vspace{-5pt} 
         \label{FIT_YBP1194}
   \end{figure}
%________________________________________________________________
%
A clear 6.9-day periodic signal can be seen in the periodogram (see
fig.\ref{FIT_YBP1194} top) with its one-year and two-year aliases on
both sides (at 6.7 d and 7.03 d).  A single-Keplerian model was
adjusted to the data (fig.\ref{FIT_YBP1194} bottom).
The resulting orbital parameters for the planet candidate are reported
in Table \ref{PlanetParam}.
The residuals' dispersion is $\sigma$(O-C)$=11.55$ \ms, comparable
with the mean measurement accuracy ($\sim$15 \ms), and the periodogram
of the residuals does not show significant power excess, although
structures are present.\\
\\
YBP1514\\
Twenty-five RV measurements have been obtained for YBP1514 since 2009:
19 with HARPS, the others with HRS and SOPHIE. The typical S/N is
$\sim$10 and the measurement uncertainty is $\sim$15 \ms for HARPS,
$\sim$25 \ms for HRS, and $\sim$10 \ms for SOPHIE.
%
%-------------------------------------------------------------
   \begin{figure}
   \vspace{-5pt}
   \centering
   \hspace{-16pt}
   \includegraphics[width=0.46\textwidth,height=3.5cm]{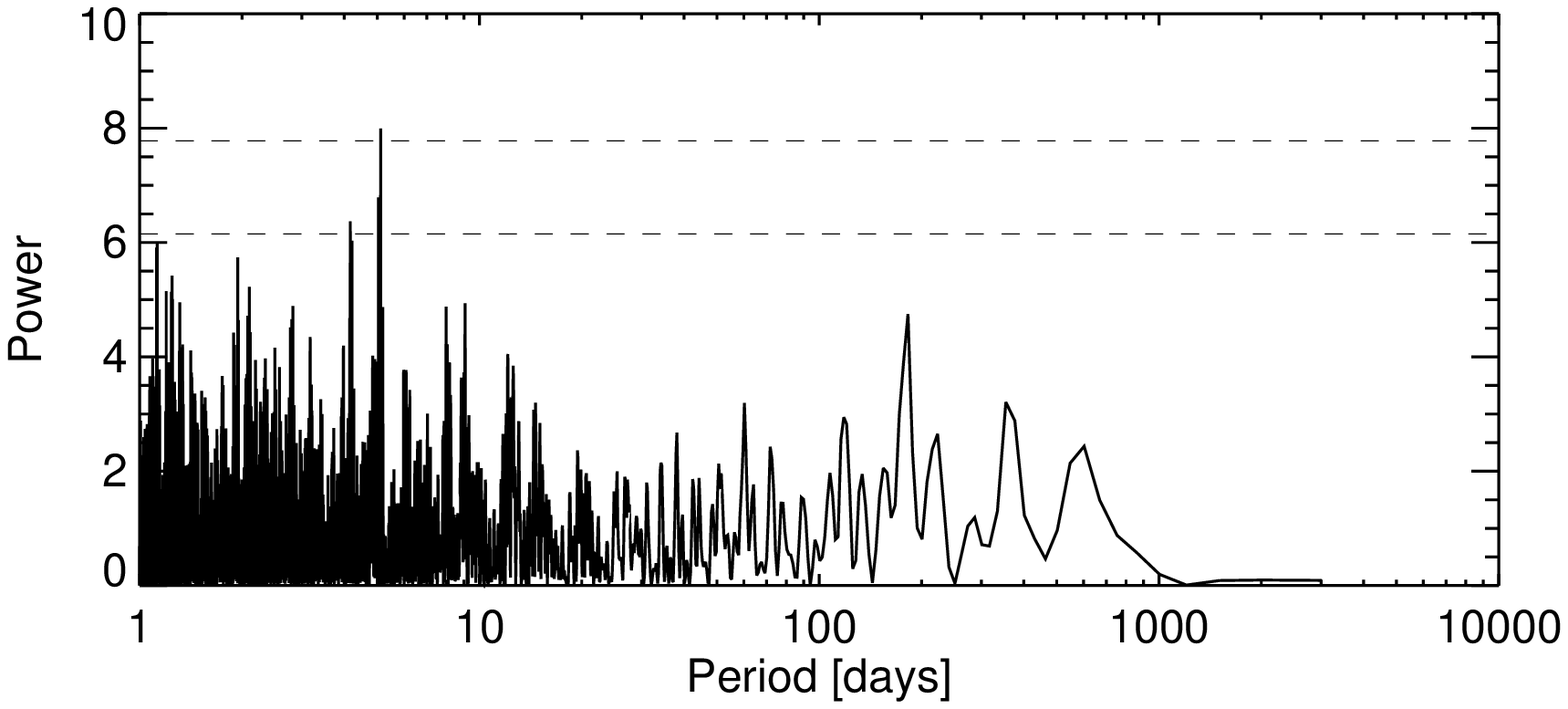}
   \includegraphics[width=0.48\textwidth]{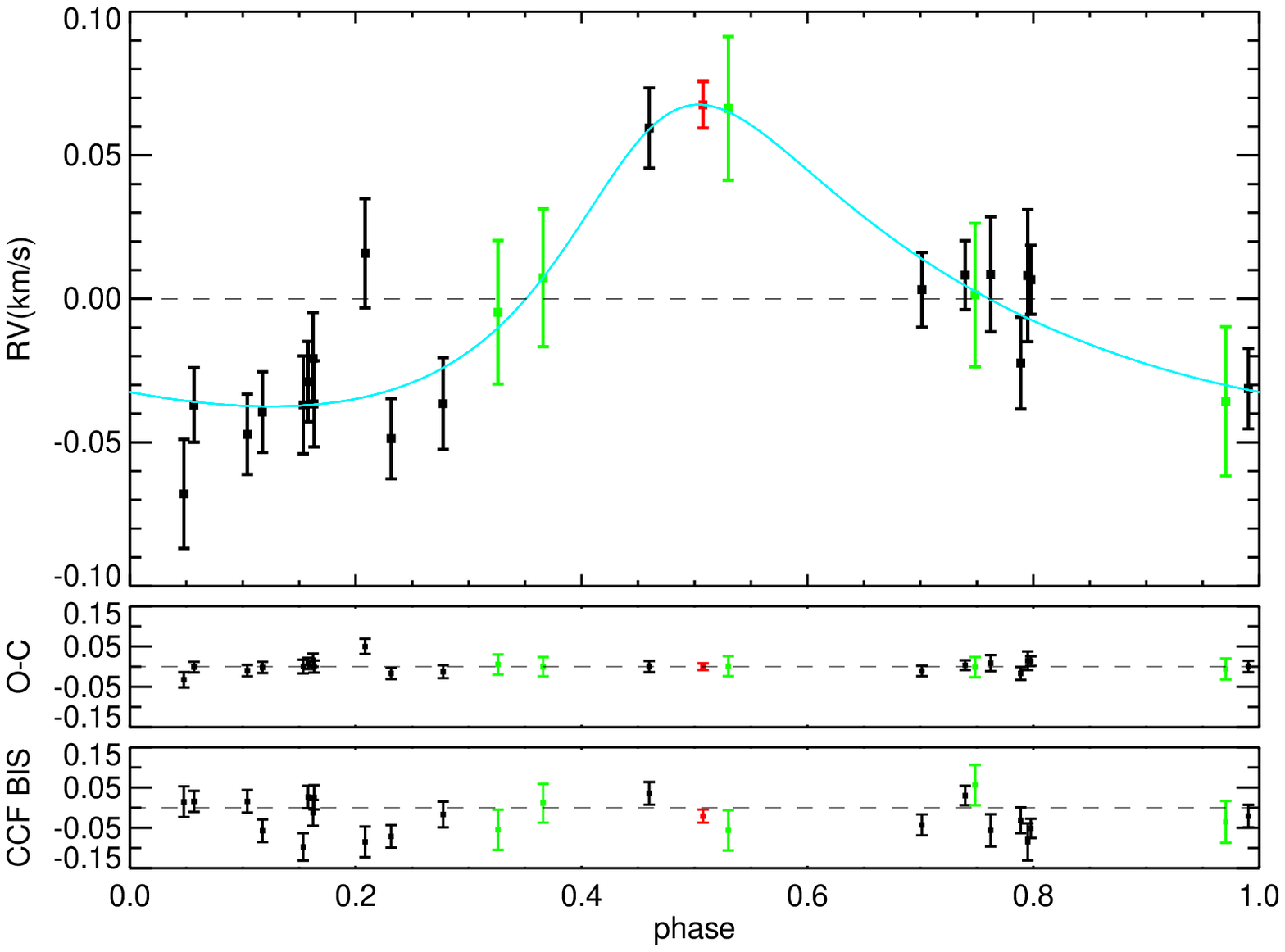}
    \vspace{-5pt}
      \caption{Top: Lomb-Scargle periodogram for YBP1514.
       Bottom: phased RV measurements and Keplerian best fit, best-fit residuals, and 
       bisector variation for YBP1514. Same symbols as in Fig.\ref{FIT_YBP1194}.
               }
         \label{FIT_YBP1514}
   \end{figure}
%
%______________________________________________________________
%
A significant peak is present in the periodogram at 5.11 days (fig.\ref{FIT_YBP1514} top),
together with its one-year alias at 5.04 days. 
We fitted a single-planet Keplerian orbit corresponding to the period
P$=5.11$ days (fig.\ref{FIT_YBP1514} bottom).  The orbital parameters
resulting from this fit are listed in Table \ref{PlanetParam}.
Assuming a mass of 0.96 M$_{\odot}$ for the host star, we computed a
minimum mass for the companion of 0.40$\pm$0.11 \Mj.
The residuals to the fitted orbit have a dispersion of
$\sigma$(O-C)$=14.6$ \ms, within the mean measurement uncertainty, and
show no significant periodicity.\\
\\
S364\\
We collected 20 radial velocity measurements of S364 in about four
years with HARPS, HRS, and SOPHIE. The average RV uncertainty is
$\sim$3.0 \ms for HARPS, $\sim$7.0 \ms for SOPHIE and $\sim$20 \ms for
HRS.  Seven additional RV measurements were obtained with CORALIE
between 2003 and 2005, with a mean measurement uncertainty of $\sim$12 \ms.
%-------------------------------------------------------------
   \begin{figure}
   \vspace{-5pt}
   \centering
   \hspace{-16pt}
   \includegraphics[width=0.46\textwidth,height=3.5cm]{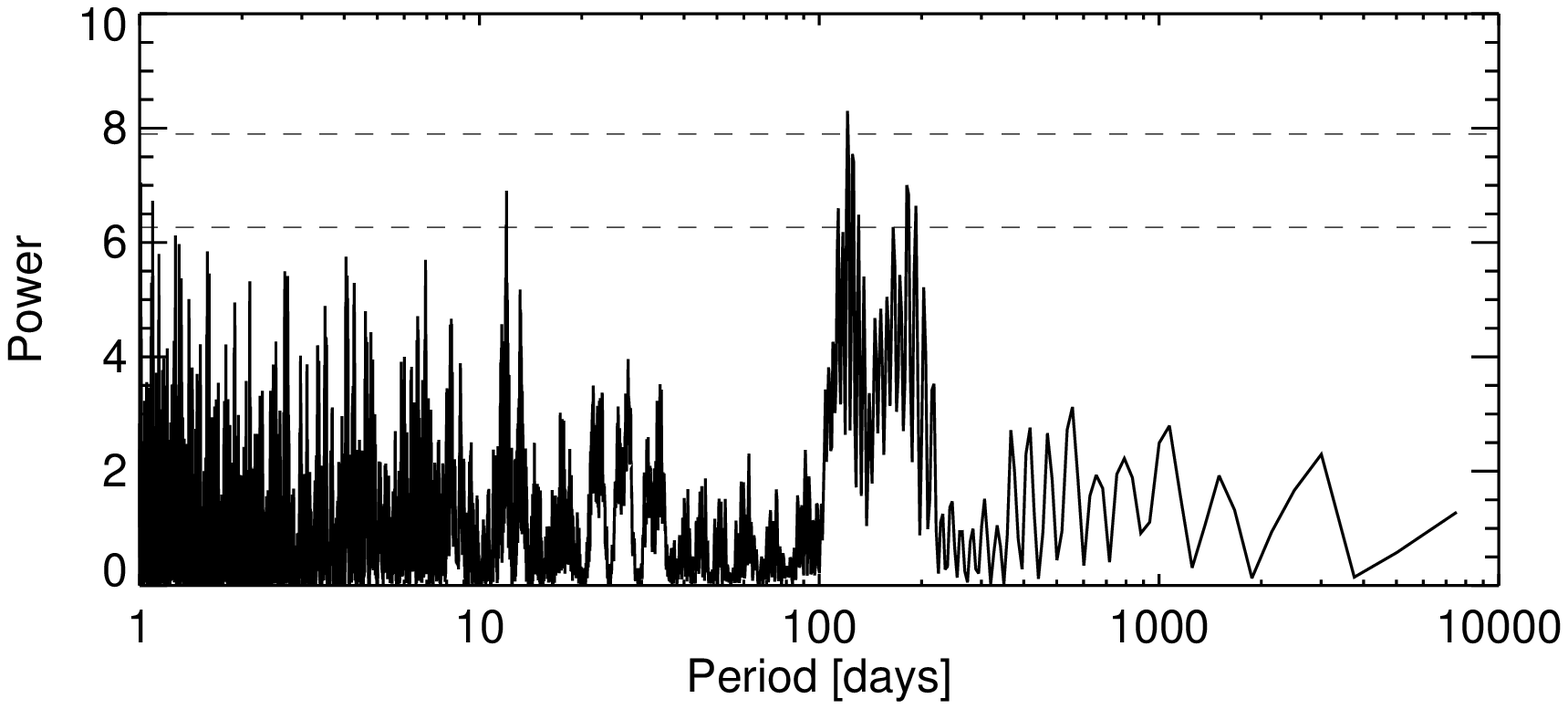}
   \includegraphics[width=0.48\textwidth]{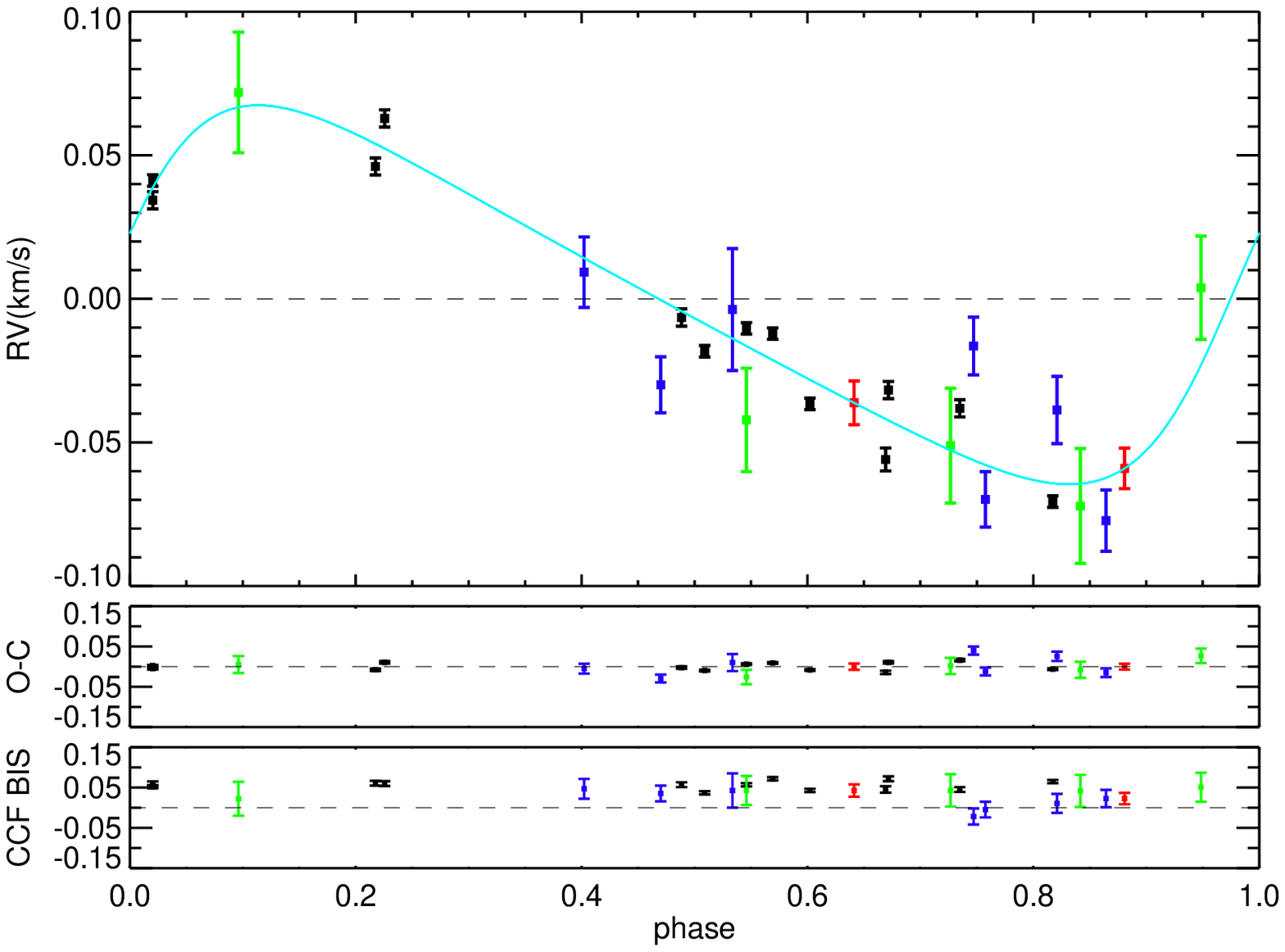}
      \vspace{-5pt}
      \caption{Top: Lomb-Scargle periodogram for S364.  Bottom: phased
        RV measurements and Keplerian best fit, best-fit residuals,
        and bisector variation for S364. Same symbols as in
        Fig.\ref{FIT_YBP1194}.  }
         \label{FIT_SAND364}
   \end{figure}
%______________________________________________________________
%
The periodogram of the observed data is shown in fig.\ref{FIT_SAND364}
(top) and indicates an excess of power at $\approx$121.7 days.  The
other clearly visible peak at 182 days is the one-year alias of the
planetary signal at P=121.7 days.  It disappears in the periodogram of
residuals, which no longer shows any signal.  We fitted a
single-planet Keplerian orbit to this signal (fig. \ref{FIT_SAND364}
bottom) and found an orbital solution whose parameters are reported in
Table \ref{PlanetParam}.
The residuals to the fitted orbit show a level of variation of
$\sigma=$16.0 \ms, higher than the estimated accuracy, but the
periodogram of the residuals does not reveal significant peaks.

\section{Discussion and prospects}
We have presented new results from our planet-search campaign in the
OC M67. Our measurements reveal that Y1194, Y1514, and S364 host
planets.

To rule out activity-related rotational modulation as the cause of the
RV variations in our object data, we investigated chromospheric
activity in these stars by measuring the variations of the core of
H$\alpha$ with respect to the continuum.  The low S/N ratio of our
observations does not provide sufficient signal in the region of the
more sensitive Ca II H and K lines.  We followed a method similar to
the one described in \citet{Pasquini1991}.  All the targets exhibit a
very low level of activity: S364 shows a variability in H$\alpha$ of
2\%, YBP1514 and YBP1194 of 3\% without significant periodicity.  In
addition, the M67 stars have a very low level of chromospheric
activity (\citet{Pace2004}: $\mathrm{<F^{'}_{K}> \sim 0.5 \cdot
  10^{6}}$ $\mathrm{erg\, cm^{-2}\, s^{-1}}$ for M67 compared with
$\mathrm{<F^{'}_{K}> \sim 2.1 \cdot 10^{6}}$ $\mathrm{erg\, cm^{-2}\,
  s^{-1}}$ for the Hyades), which is not compatible with generating
the high RV variations we observe. Therefore, rotationally modulated
RV variations for the dwarfs in M67 are certainly not a concern.  The
remote possibility that these stars are short-period binaries seen
pole-on can also be excluded, because they are very active, and will
show enhanced H$\alpha$ cores and strong X-ray emission, which has not
been observed for these stars \citep{VandenBerg2004}.  The fact that
these stars are of solar age and that our research is focused on
finding giant planets with an expected RV variability of tens of \ms
makes the contamination by activity irrelevant.

It is remarkable that Y1194 is one of the best-known solar twins.
This star together with Y1514, S364, and the other M67 targets will be
suitable for a detailed differential abundance analysis to compare the
chemical composition of stars with and without giant planets.\\
All the orbital solutions show nonzero eccentricity, but this is also
common among planets found around field stars.
\citet{Quinn2013} explained that hot-Jupiters in OCs with nonzero
eccentric orbits and circularization time-scales $t_{circ}$ longer
than the system age, might provide an observational signature of the
hot-Jupiter migration process via planet-planet scattering.  We
evaluated $t_{circ}$ for the eccentric orbits of YBP1194 and YBP1514.
Assuming a tidal quality factor $6\times10^{4}<Q_{p}<2\times10^{6}$,
we calculated $409 Myr<t_{circ}<13.6 Gyr$ for YBP1194 and $220
Myr<t_{circ}<6.9 Gyr$ for YBP1514 (see \citet{Quinn2013} for details).
Given the solar age of M67 and the wide range of possible $t_{circ}$,
reflecting the choice of the $Q_{p}$ and the estimation of the
planetary radius, no firm conclusion can be drawn for the origin of
the eccentric short-period orbits of these stars.  Moreover, further
investigations and more RV data are needed to better constrain the
eccentricities of these objects \citep[see][]{Pont2011}.\\
The planet around the giant S364 belongs to the low-populated region
of periods between $\sim$10 and $\sim$200 days and is one of the
shortest periods found around evolved stars.

When we examine the current distribution of the Jupiter-mass planets
for RV surveys around FGK stars we find an exoplanet host-rate higher
than 10\% for planets with a period of up to a few years and
1.20$\pm$0.38\% at solar metallicity, for very close-in hot-Jupiters
with a period shorter than ten days \citep{Cumming2008, Mayor2011, Wright2012}.
This rate around field stars has been in contrast to the lack of
detected planets in both open and globular cluster for several years.
Before 2012, the detections were limited to a long-period giant planet
around one of the Hyades clump giants \citep{Sato2007} and to a
substellar-mass object in NGC2423 \citep{Lovis2007}.
No evidence of short-period giant planets has been presented in the
study of \citet{Paulson2004} around main-sequence stars of the Hyades,
or in several transit campaigns
\citep{Bramich2005,Mochejska2005,Mochejska2006,Pepper2008,Hartman2009}.
These triggered the hypothesis that the frequency of planet-hosting
stars in clusters is lower than in the field.  To explain the
dichotomy between field and cluster stars, it has been suggested that
the cluster environment might have a significant impact on the
disk-mass distribution.  \citet{Eisner2008}, studying disks around
stars in the Orion Nebula Cluster (ONC), proposed that most of these
stars do not posses sufficient mass in the disk to form Jupiter-mass
planets or to support an eventual inward migration. Other scenarios
may be attributed to post-formation dynamics, in particular to the
influence of close stellar encounters \citep{Spurzem2009, Bonnell2001}
or to tidal evolution of the hot-Jupiters \citep{Debes2010} in the
dense cluster environment.  \citet{VanSaders2011}, in contrast, found
no evidence in support of a fundamental difference in the short-period
planet population between clusters and field stars, and attributed the
nondetection of planets in transit surveys to the inadequate number of
stars surveyed.  This seems to be confirmed by the recent results.
Indeed, we can list the discovery of two hot-Jupiters in the Praesepe
open cluster in 2012 \citep{Quinn2012} and of two sub-Neptune planets
in the cluster NGC6811 as part of The Kepler Cluster Study
\citep{Meibom2013}, the new announcement of a hot-Jupiter in the
Hyades \citep{Quinn2013} and now the detection in M67 of three
Jupiter-mass planets presented in this work.
\citet{Quinn2012} obtained a lower limit on the hot-Jupiter frequency
in Praesepe of 3.8$^{+5.0}_{-2.4}$\%, which is consistent with that of
field stars considering the enriched metallicity of this cluster.
\citet{Meibom2013} have found the same properties and frequency of
low-mass planets in open clusters as around field stars.  In our case,
for short-period giant planets we derived a frequency of
2$^{+3.0}_{-1.5}$\% (errors computed according to
\citet{Gehrels1986}); which is slightly higher than the value for
field stars.  Adding giant planets with long periods, the rate becomes
3.4$^{+3.3}_{-1.7}$\%, but this fraction is a lower limit that will
increase with the follow-up of some other candidates
\citep[see][]{Pasquini2012}, which reveal suggestive signals for
additional planetary companions.  If these were confirmed,
the  frequency of giant planets would rise to 13$^{+5.0}_{-2.5}$\%, in agreement with the rate of giant planets
found by \citet{Mayor2011} for field stars.
%%
%%%
%%
\begin{acknowledgements}
LP acknowledges the Visiting Researcher program of the CNPq Brazilian
Agency, at the Fed. Univ. of Rio Grande do Norte, Brazil.  RPS thanks
ESO DGDF, the Hobby Eberly Telescope (HET) project, the PNPS and PNP
of INSU - CNRS for allocating the observations.  MTR received support
from PFB06 CATA (CONICYT).  We are grateful to Gaspare Lo Curto and
Michele Cappetta for the support in data reduction analysis and
helpful discussions.
\end{acknowledgements}

%-------------------------------------------------------------------

%-------------------------------------------------------------------------------
%
\appendix 
\Online

\section{CMD and membership probabilities}

In this section we summarize the results presented in
\citet{Pasquini2012}, focusing in particular on the three stars
discussed in the letter.\\ YBP1194, YBP1514, and S364 belong to the
M67 sample that includes a total of 88 stars.  All targets have V
mag. between 9 and 15, and a mass range between 0.9-1.4 M$_{\odot}$.\\

%
%                 A figure as large as the width of the column
%-------------------------------------------------------------
   \begin{figure}[!h]
   \vspace{-5pt}
   \centering
   \hspace{-16pt}
   \includegraphics[width=0.48\textwidth]{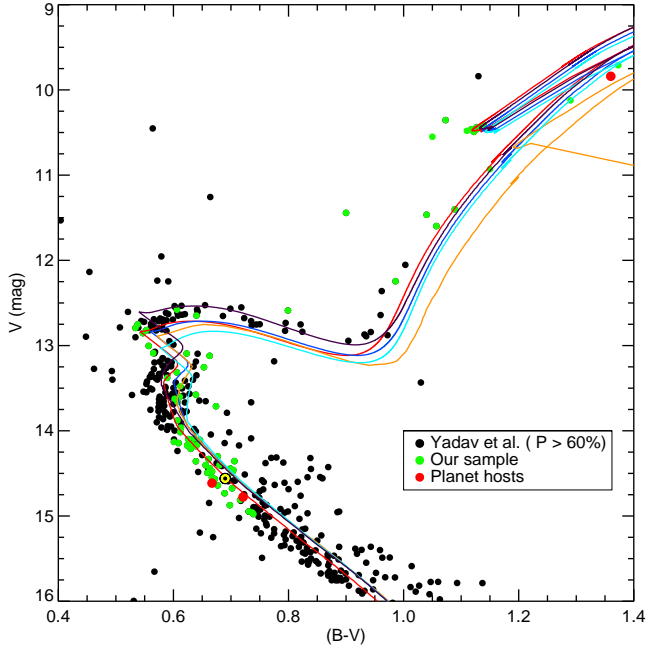}
      \vspace{-5pt}
      \caption{CMD of M67 \citep[photometry from][]{Yadav2008} for probable members (P$_{\mu}>60\%$).
              The isochrones are taken from the BaSTI library \citep{Pietrinferni2004}. 
              The isochrones in black, dark blue, and light blue correspond to 3.5 Gyr, 4.0 Gyr, and 4.5 Gyr with 
              a reddening E(B-V)=0.041$\pm$0.004 \citep{Taylor2007}.
              The isochrone in red is a 4.0 Gyr with a lower reddening (E(B-V)=0.02).
              The isochrone in orange is a 4.47 Gyr from \citet{Girardi2000} with E(B-V)=0.041$\pm$0.004.  
              %such as used in \citet{Pasquini2012}.
              The location of the Sun, if it were within M67, is marked with a $\odot$ in yellow.
              }
         \label{DiagHR_Candidates}
   \end{figure}
%-------------------------------------------------------------
%

We selected main-sequence stars (included YBP1194 and YBP1514) with a
membership probability higher than 60\% and a proper motion shorter
than 6 mas/yr with respect to the average according to
\citet{Yadav2008}.  For the giants we refer to \citet{Sanders1977}.
The RV membership was established for the latter following the work of
\citet{Mermilliod2007}, who studied the membership and binarity of 123
red giants in six old open clusters, and of \citet{Mathieu1986}, who
made a very complete RV survey of the evolved stars of M67 with a
precision of a few hundreds of \ms.  The majority of the other stars
were selected according to \citet{Pasquini2008}, who used several
VLT-FLAMES exposures for each star to classify suspected binaries.  We
found that YBP1194, YBP1514, and S364 are probable RV members with a
mean radial velocity within one-sigma from the average cluster RV.
For the latter, we adopted the value of $\langle RV_{M67}\rangle =
33.724$\kms and the dispersion of $\sigma = \pm0.646$\kms estimated in
\citet{Pasquini2012}.\\
Table 3 shows proper motions and membership probability for the three stars discussed.
\begin{table}
\caption{Object ID, proper motions, and
 membership probability of the targets; reference.}
\label{Membership}
\centering
\small
\vspace{-5pt}
\begin{tabular}{lcccc}
\hline\hline
%\hline
Object        & $\mathrm{\mu_{x}\pm\Delta \mu_{x}}$
                                & $\mathrm{\mu_{y}\pm\Delta \mu_{y}}$ 
                                                  & Prob$\%$ 
                                                       & Reference                      \\
\hline
YBP1194       &  0.30$\pm$1.01  & -0.42$\pm$0.65  & 99 & \textrm{Yadav et al.\ (2008)}    \\
YBP1514       & -0.12$\pm$1.13  &  1.73$\pm$1.37  & 98 & \textrm{Yadav et al.\ (2008)}    \\
S364          & -0.088          &  0.164          & 82 & \textrm{Yadav et al.\ (2008)}    \\ 
\hline
\end{tabular}
\normalsize
\end{table}

Details about selection criteria and motion errors can be found in the
original \citet{Yadav2008} and \citet{Sanders1977} works.\\ In Figure
\ref{DiagHR_Candidates}, we report the observed region of the
color-magnitude diagram (CMD), indicating in different colors the the
position of the stars considered in this letter and the solar analog,
as determined in \citet{Pasquini2008}.  The three stars analyzed in
this work lie quite well on the cluster sequence in the CMD.  We
superimposed the isochrones from \citet{Pietrinferni2004} with solar
metallicity and age corresponding to 3.5 Gyr (black curve), 4.0 Gyr
(dark-blue curve) and 4.5 Gyr (light-blue curve).  We also included
the 4.0 Gyr isochrone (red curve) with a slightly lower reddening
(E(B-V)=0.02 instead of 0.041 (Taylor 2007)).
This curve seems to match the colors of the turnoff better \citep[see
  also the discussion in][]{Pasquini2012}.  In the same figure, we
report the Padova isochrone using E(B-V)=0.041$\pm$0.004, with solar
metallicity, age 4.47 Gyr, and Y=0.26 \citep{Girardi2000}.\\ Given
that the values of stellar parameters have influence on the estimation
of the planet masses, we evaluated the effects on the host star masses
and radii of using isochrones with different ages and slightly lower
reddening.  While for the two main-sequence stars YBP1194 and YBP1514
we found no significant incidence, for the giant S364, an age
uncertainty of $\pm$0.5 Gyr and a lower reddening would induce an
error on the star mass of 4$\%$ and on its radii of 3$\%$.  Therefore,
we decided to include this effect in the uncertainties of S364 listed
in Table \ref{StarParam} and in the error of the planet mass.
 
 %_____________________________________________________________
%
\begin{table*}
\caption{Relative RV measurements, RV uncertainties, bisector span,
  and ratio of the H$_{\alpha}$ core with respect to the continuum
  \citep[see][]{Pasquini1991} for YBP1194.  All the RV data points are
  corrected to the zero point of HARPS.}
\label{table_YBP1194}      
\centering          
\begin{tabular}{lrrrrl }     % 5 columns 
\hline\hline        
                      % To combine 4 columns into a single one 
  BJD & RV & $\sigma RV$ & BIS span& H$_{\alpha}$ratio& instrument\\
  (-2450000) & (\kms) & (\kms)& (\kms) &  & \\
\hline                    
  4489.51193	&$-$0.000	&0.009     &$-$0.021029 & 0.038294    & Sophie \\%
  4491.50617	&$-$0.030	&0.009     &$-$0.042000 & 0.038689    & Sophie \\%
  4842.84025	&   0.051	&0.010     &$-$0.018851 & 0.038897     & Harps \\%
  4856.62544	&   0.013	&0.012     &$-$0.081895 & 0.038995    & Harps \\%
  4862.59495	&   0.031	&0.013     &$-$0.009653 & 0.039465    & Harps \\ 
  5188.83049	&   0.002	&0.010     &$-$0.056037 & 0.039817    & Harps \\
  5189.82037	&   0.010	&0.010     &   0.022874 & 0.039387    & Harps  \\
  5190.79901	&   0.036	&0.009     &   0.039774 & 0.038994    & Harps \\
  5214.85851	&$-$0.030	&0.013     &   0.008036 & 0.039872    & Harps \\
  5216.70466	&$-$0.000	&0.012     &   0.051777 & 0.037107    & Harps \\
  5594.79168	&   0.019	&0.014     &$-$0.021881 & 0.039218    & Harps \\
  5977.66236	&   0.037	&0.010     &$-$0.028000 & 0.037126    & Harps \\
  5986.51471	&$-$0.034	&0.009     &$-$0.057833 & 0.038907    & Sophie \\%
  6219.98852	&   0.049	&0.027     &$-$0.025204 & 0.037364    & Het  \\
  6243.93406	&$-$0.028	&0.028     &   0.019138 & 0.036729    & Het  \\
  6245.81040	&$-$0.007	&0.017     &$-$0.026488 & 0.041533    & Harps \\
  6270.77262	&   0.009	&0.019     &$-$0.053592 & 0.039405    & Harps  \\
  6286.00446	&$-$0.066	&0.025     &$-$0.042748 & 0.036126    & Het    \\
  6305.17913	&   0.017	&0.014     &$-$0.092977 & 0.038956    & Harps  \\
  6316.76841	&   0.005	&0.012     &$-$0.021029 & 0.039405    & Harps  \\%
  6322.71086	&$-$0.006	&0.021     &$-$0.028423 & 0.038896    & Harps  \\%
  6338.66079	&   0.034	&0.026     &$-$0.095893 & 0.037364    & Het    \\
  6378.72343	&   0.002	&0.026     &   0.024230 & 0.038289    & Het    \\
\hline                  
\end{tabular}
\end{table*}
%
%_____________________________________________________________
%
\begin{table*}
\caption{Relative RV measurements, RV uncertainties, bisector span, and ratio of the H$_{\alpha}$ core 
  with respect to the continuum \citep[see][]{Pasquini1991} for YBP1514. 
  All the RV data points are corrected to the zero point of HARPS.}             
\label{table_YBP1514}      
\centering          
\begin{tabular}{lrrrrl }     % 5 columns 
\hline\hline        
                      % To combine 4 columns into a single one 
  BJD & RV & $\sigma RV$ & BIS span& H$_{\alpha}$ratio& instrument\\
 (-2450000) & (\kms) & (\kms)& (\kms) &  &\\
 \hline
 4858.72562	& $-$0.037  & 0.017  & $-$0.097357  & 0.040203 & Harps\\
 4861.71515	&    0.008  & 0.012  &    0.030042  & 0.039426 & Harps\\
 5214.87795	&    0.008  & 0.020  & $-$0.056154  & 0.037161 & Harps\\
 5216.72426	& $-$0.047  & 0.014  &    0.015647  & 0.042089 & Harps\\
 5260.70288	&    0.003  & 0.013  & $-$0.042552  & 0.039905 & Harps\\
 5269.72337	&    0.059  & 0.014  &    0.035491  & 0.038077 & Harps\\
 5595.74177	& $-$0.029  & 0.014  &    0.026610  & 0.039282 & Harps\\
 5626.15072	& $-$0.039  & 0.014  & $-$0.057171  & 0.039905 & Harps\\ %
 5943.29835	& $-$0.037  & 0.013  &    0.015647  & 0.040103 & Harps\\ %
 5967.45833	&    0.008  & 0.023  & $-$0.084845  & 0.039356 & Harps\\ %
 5968.59423	& $-$0.031  & 0.014  & $-$0.020833  & 0.039036 & Harps\\ %
 5977.68405	&    0.007  & 0.012  & $-$0.051313  & 0.039475 & Harps\\
 5986.56139	&    0.068  & 0.010  & $-$0.020833  & 0.039306 &  Sophie\\ %
 6036.66109	& $-$0.005  & 0.025  & $-$0.054711  & 0.040201 &  Het \\
 6245.83389	& $-$0.037  & 0.015  &    0.025824  & 0.040109 &  Harps\\
 6254.90496	& $-$0.036  & 0.026  & $-$0.035278  & 0.040001 &  Het\\
 6270.81399	& $-$0.068  & 0.019  &    0.014902  & 0.039345 &  Harps \\
 6305.22499	& $-$0.022  & 0.016  & $-$0.031263  & 0.037032 &  Harps\\
 6307.81028	& $-$0.036  & 0.016  & $-$0.016596  & 0.039764 &  Harps\\ %
 6317.74917	& $-$0.049  & 0.014  & $-$0.071034  & 0.038687 &  Harps \\ %
 6322.68946	&    0.016  & 0.019  & $-$0.084845  & 0.040020 &  Harps \\ %
 6332.68338	& $-$0.021  & 0.016  & $-$0.012640  & 0.039854 &  Harps \\
 6335.68811	&    0.001  & 0.025  &    0.056131  & 0.041020 & Het \\
 6339.68307	&    0.066  & 0.025  & $-$0.056093  & 0.042437 & Het  \\
 6364.59175	&    0.007  & 0.024  &    0.011085  & 0.040285 & Het \\
\hline                  
\end{tabular}
\end{table*}
%
%_____________________________________________________________
 %
 \begin{table*}
 \caption{Relative RV measurements, RV uncertainties, bisector span, and ratio of the H$_{\alpha}$ core 
  with respect to the continuum \citep[see][]{Pasquini1991} for S364.
  All the RV data points are corrected to the zero point of HARPS.}             
 \label{table_S364}      
 \centering          
 \begin{tabular}{lrrrrl }     % 5 columns 
 \hline\hline        
                       % To combine 4 columns into a single one 
   BJD & RV & $\sigma RV$ & BIS span& H$_{\alpha}$ratio& instrument\\
   (-2450000) & (\kms) & (\kms)& (\kms) &  &\\
  \hline 
  2647.77191	&  $-$0.030   &  0.010 &    0.035310&      -     &  Coralie\\
  2682.68790	&  $-$0.070   &  0.010 & $-$0.004670&      -     &  Coralie\\ 
  2695.70992	&  $-$0.077   &  0.011 &    0.022800&      -     &  Coralie\\
  3004.82521	&     0.009   &  0.012 &    0.046980&      -     &  Coralie\\
  3020.78490	&  $-$0.004   &  0.021 &    0.042690&      -     &  Coralie\\
  3046.73656	&  $-$0.016   &  0.010 & $-$0.021670&      -     &  Coralie\\
  3055.71453	&  $-$0.039   &  0.012 &    0.010630&      -     &  Coralie\\
  4855.58011	&  $-$0.037   &  0.002 &    0.042682&  0.038161  &  Harps\\
  4860.32500	&  $-$0.036   &  0.008 &    0.042690&  0.038048  &  Sophie\\
  5216.82809	&  $-$0.012   &  0.002 &    0.071695&  0.038278  &  Harps\\
  5594.58582	&  $-$0.032   &  0.003 &    0.071762&  0.038105  &  Harps\\
  5977.55818	&  $-$0.071   &  0.002 &    0.064832&  0.037733  &  Harps\\
  5985.30291	&  $-$0.059   &  0.007 &    0.022666&  0.038426  &  Sophie\\ %    
  6236.94284	&     0.004   &  0.018 &    0.050689&  0.040027  &  Het\\  
  6245.76368	&     0.034   &  0.003 &    0.059181&  0.038501  &  Harps\\
  6245.86577	&     0.041   &  0.002 &    0.052050&  0.038271  &  Harps\\
  6269.71823	&     0.046   &  0.003 &    0.061007&  0.041626  &  Harps\\
  6270.75931	&     0.063   &  0.003 &    0.059963&  0.038616  &  Harps\\
  6302.68302	&  $-$0.006   &  0.003 &    0.057069&  0.038056  &  Harps\\
  6305.24544	&  $-$0.018   &  0.002 &    0.036772&  0.037843  &   Harps\\
  6309.75038	&  $-$0.010   &  0.002 &    0.056751&  0.039626  &  Harps\\%
  6309.75141	&  $-$0.042   &  0.018 &    0.042921&  0.039879  &  Het\\ 
  6324.67965	&  $-$0.056   &  0.004 &    0.045385&  0.038311  &  Harps\\%
  6331.69998	&  $-$0.051   &  0.020 &    0.043008&  0.039356  &  Het\\ 
  6332.69784	&  $-$0.038   &  0.003 &    0.045065&  0.038067  &  Harps\\%
  6345.65885	&  $-$0.072   &  0.020 &    0.041555&  0.039587  &  Het\\ 
  6376.73341	&     0.072   &  0.021 &    0.022076&  0.039445  &  Het\\
 \hline       
 \end{tabular}
\end{table*}
 
%-------------------------------------------------------------
\end{document}